\documentclass[ nofootinbib, showpacs,
amsmath, amssymb]{revtex4}

\begin{document}

\title{ Minimum-length deformed quantization of a free field on de Sitter background and corrections to the inflaton perturbations}

\author{ Michael~Maziashvili}
\email{maziashvili@gmail.com} \affiliation{Center for Elementary Particle Physics, ITP, Ilia State University, 3/5 Cholokashvili Ave., Tbilisi 0162, Georgia }

\begin{abstract}

The effect of string and quantum gravity inspired minimum-length deformed quantization on a free, massless scalar field is studied on de Sitter background at the level of second quantization. Analytic solution of a field operator is obtained to the first order in deformation parameter. Using this solution we then estimate the two point and four point correlation functions (with respect to the Bunch-Davies vacuum). The field operator shows up a non-linear dependence on creation and annihilation operators, therefore the perturbation spectrum proves to be non-Gaussian. The correction to the power spectrum is of the same order as obtained previously in a similar study that incorporates the minimum-length deformed momentum operator into the first quantization picture and then proceeds in the standard way for second quantization. The non-Gaussianity comes at the level of four point correlation function; its magnitude appears to be suppressed by the factor $\sim\exp(-6N)$, where $N$ is the number of e-foldings.

\end{abstract}

\pacs{03.70.+k; 04.60.-m; 04.60.Bc; 98.80.Cq }





\maketitle

\section{ Introduction  }
\label{Introduction}

A striking consequence of inflationary picture is that all structures we observe today owe their existence to the very small quantum fluctuations (of Planck length and even smaller) occurring during the inflationary epoch. So, it represents a very instructive example of the strong connection between Planck scale physics and the large-scale universe \cite{Martin:2000xs, Brandenberger:2000wr}. In view of the high precision cosmological measurements the study of possible imprints of quantum gravity on the cosmological perturbations could open an experimental window for testing of Planck scale physics. The aim of this paper is to study further the influence of minimum-length deformed quantum mechanics
\begin{equation} [Q,\,P] = i (1 \,+\, \beta P^2)~, \label{mlqmonedim}\end{equation}
upon the inflaton perturbation spectrum. (We will assume natural system of units $\hbar = c = 1$ throughout this paper.) In previous studies pertaining to the cosmological perturbations \cite{Kempf:2000ac, Kempf:2001fa, Niemeyer:2001qe, Niemeyer:2002ze, Hassan:2002qk, Ashoorioon:2004vm, Palma:2008tx} most of the results were obtained by incorporating the minimum-length deformed momentum operator in the first quantization picture and then the standard scheme for a second quantization was applied. To date, in the framework of this approach a substantial amount of theoretical work has been devoted to various phenomenological questions. The following obvious step is to complete this study by undertaking the minimum-length deformed quantization at the level of second quantization. A Specific objective of this paper is the quantization a massless scalar field on the de Sitter background with respect to the minimum-length deformed formalism, Eq.\eqref{mlqmonedim}. For this purpose we shall apply the theoretical framework for second quantization with respect to the minimum-length deformed quantum mechanics put forward in \cite{Mania:2009dy, Berger:2010pj}. First we write down the corrected Hamiltonian, then solve Heisenberg equation of motion for field operator to the first order in $\beta$, and then estimate two point and four point correlation functions with respect to the Bunch-Davies vacuum.

Let us now somewhat quantify our discussion. The minimum-length deformed quantum mechanics Eq.\eqref{mlqmonedim} underlies the generalized uncertainty relation 

\begin{equation}  \delta Q \delta P  \geq  \frac{1}{2} \left( 1 + \beta \delta P^2 \right) ~, \label{gur}\end{equation} proposed originally in the context of perturbative string theory as a consequence of the fact that strings cannot probe distances below the string scale (string length) \cite{String}. This relation immediately imposes lower bound on position uncertainty $\delta Q_{min} = \sqrt{\beta}$. Further, Eq.\eqref{gur} was discussed by combining the basic principles of quantum theory and general relativity in the framework of {\tt Gedankenexperiments} \cite{heuristic}. The parameter $\beta$ is set by the quantum gravity scale (either by string length or Planck length which are of the same order) $\sqrt{\beta} \sim l_P \simeq 10^{-33}$\,cm. The deformed position and momentum operators in Eq.\eqref{mlqmonedim} can be solved in terms of the standard $q,\,p$ operators $[q,\,p]=i$ (see the introduction of paper \cite{Kempf:1996nk}) 

\begin{equation}\label{onedimexactmodops} Q = q,~~P = \beta^{-1/2}\tan\left(p\sqrt{\beta}\right)~. \end{equation} In what follows we will restrict our consideration to the first order in $\beta$, that is, we will use the following truncation  

\begin{equation}\label{kempfmangano} Q = q,~~P = p + \frac{\beta}{3}\,p^3 ~. \end{equation}

\noindent The multidimensional generalization of Eq.\eqref{mlqmonedim} that preserves translation and rotation invariance and introduces a finite minimum position uncertainty in all three position variables has the form \cite{Kempf:1996fz, Kempf:2000ac, Ashoorioon:2004vm}

\begin{equation} \left[Q^i,\,P^j\right] = i\left( \frac{2\beta \mathbf{P}^2}{\sqrt{1+4\beta \mathbf{P}^2} \,-\, 1}\,\delta^{ij} +2\beta P^iP^j  \right)\,,~~ \left[Q^i,\,Q^j\right] = \left[P^i,\,P^j\right]=0~.\label{minlengthqm} \nonumber\end{equation} The deformed $\mathbf{Q},\,\mathbf{P}$ operators can be represented in terms of the standard $\mathbf{q},\,\mathbf{p}$ operators in the following way 

\begin{equation}\label{deformedopmultid} Q^i = q^i\,,~~~~P^i = \frac{p^i}{1-\beta \mathbf{p}^2}~. \end{equation} Its Hilbert space representation (in standard-momentum $\mathbf{p}$ representation) has the form \[ Q^i\psi(\mathbf{p}) = i\partial_{p_i} \psi(\mathbf{p})\,,~~~~ P^i\psi(\mathbf{p}) = \frac{p^i}{1-\beta \mathbf{p}^2}\,\psi(\mathbf{p})~, \]  with the scalar product 

\[ \langle \psi_1 | \psi_2 \rangle = \int\limits_{\mathbf{p}^2 < \beta^{-1}}d^3p \,\psi^*_1(\mathbf{p})\psi_2(\mathbf{p})~. \] To the linear order in $\beta$ Eq.\eqref{deformedopmultid} reads

\begin{equation} Q_i = q_i~,~~~~ P_i = p_i \left[ 1 + {\beta} \left(\mathbf{p}\right)^2 \right] ~.\label{firstorderapproxmodop}\end{equation} In the classical limit the Eq.\eqref{firstorderapproxmodop} yields the modified dispersion relation 

\begin{equation} \label{moddisperrel1}  \varepsilon^2 = {\mathbf p}^2 + m^2 + 2{\beta} {\mathbf p}^4~. \end{equation} admiting simple physical interpretation that due to quantum-gravitational fluctuations of the background metric, the energy $ \varepsilon = \sqrt{{\mathbf p}^2 + m^2}$ acquires the increment ${\beta} {\mathbf p}^4/\sqrt{{\mathbf p}^2 + m^2}$.

The corrections to the field theory due to minimum-length deformed quantum mechanics comes in two parts. 

{\bf 1. Modification at the first quantization level.} At the first quantization level the modification amounts to the replacement of momentum operator in field Lagrangian with the deformed operator, Eq.\eqref{firstorderapproxmodop}. That is, the modified field theory takes the form \cite{Kempf:1996nk, Kempf}

\begin{equation}\label{scaction}  \mathcal{A}[\varPhi] = - \int d^4x \, \frac{1}{2} \left[\varPhi\partial_t^2\Phi  + \varPhi{\mathbf P}^2\varPhi 
+ m^2\varPhi^2 \right]~,\end{equation} where $\mathbf{P}$ denotes deformed momentum, Eq.\eqref{firstorderapproxmodop} (just to restrict ourselves for simplicity to the leading order in $\beta$). The action
 \eqref{scaction} results in the equation of motion 

\[ \partial_t^2\Phi -
\bigtriangleup\varPhi +  2\beta \bigtriangleup\bigtriangleup\varPhi + m^2\varPhi = 0 ~,
\] the plan wave solutions for which can be obtained by the replacement 

\begin{eqnarray} \exp\left(i\left[\mathbf{p}\mathbf{x} - \sqrt{{\mathbf p}^2 + m^2}\, t \right]\right) ~~\rightarrow  ~~ \exp\left(i\left[\mathbf{p}\mathbf{x} - \left(\sqrt{{\mathbf p}^2 + m^2} + \frac{{\beta} {\mathbf p}^4}{\sqrt{{\mathbf p}^2 + m^2}}\right) t \right]\right) ~. \label{dispwave}\end{eqnarray} The equation \eqref{dispwave} explicitly reflects the modified dispersion relation \eqref{moddisperrel1}. In addition to this modification, the theory given by the action functional \eqref{scaction} involves the cutoff $\mathbf{p}^2 <\beta^{-1}$ (that is evident from Eqs.(\ref{onedimexactmodops}, \ref{deformedopmultid})). Once the Lagrangian is specified, one then proceeds in the standard way for the second quantization.

{\bf 2. Modification at the second quantization level.} The second related departure from the usual quantum field theory takes place when we are quantizing the field with respect to the deformed quantization \cite{Mania:2009dy, Berger:2010pj}. Let us ignore the modification of the field theory at the first quantization level, Eq.\eqref{scaction}, and consider second quantization with respect to the minimum-length deformed prescription. Assume $l$ is an intrinsic length scale to the problem under consideration. After introducing of ladder operators the Hamiltonian of a free field takes the form \cite{Bogolyubov:1980nc}

\begin{eqnarray}\label{desctherulseforsmlq} \mathcal{H} &=& \frac{1}{2}  \int d^3 k \, \omega_{\mathbf{k}}  \left[ b^+(\mathbf{k})b(\mathbf{k}) + b(\mathbf{k})b^+(\mathbf{k}) \right]   ~, \end{eqnarray} where $\omega_{\mathbf{k}} = \sqrt{\mathbf{k}^2}$ (massless field is considered). Defining real variables 

\begin{eqnarray}  Q_{\mathbf{k}} =\, \sqrt{\frac{l^{-2}}{  2 \omega_{\mathbf{k}} }} \left[ b(\mathbf{k}) + b^+(\mathbf{k})\right]~,~~ P_{\mathbf{k}} = \, i  \sqrt{\frac{ \omega_{\mathbf{k}}l^{-4}}{2}} \left[b^+(\mathbf{k}) - b(\mathbf{k})\right] ~,\nonumber \end{eqnarray} the Hamiltonian \eqref{desctherulseforsmlq} splits into a sum of independent one-dimensional oscillators \cite{Bogolyubov:1980nc}

\begin{equation}\mathcal{H} =   \int \frac{d^3k}{l^{-3}} \, \left( \frac{ P_{\mathbf{k}}^2}{2l^{-1}} \,+\, \frac{l^{-1}\omega_{\mathbf{k}}^2 Q_{\mathbf{k}}^2}{2} \right)~,\label{modhambm}\end{equation} each heaving the mass $l^{-1}$. While the appearance of the length scale $l$ in Eq.\eqref{modhambm} is of no importance in the framework of standard quantization (because the energy spectrum of harmonic oscillator does not depend on its mass), it appears explicitly in the energy spectrum when the minimum length deformed quantization is applied \cite{Kempf:1996fz}. Namely, assuming $P_{\mathbf{k}},\, Q_{\mathbf{k}}$ are minimum-length deformed operators and using Eq.\eqref{kempfmangano}, the Hamiltonian \eqref{modhambm} to the first order in $\beta$ takes the form 

\begin{eqnarray}  \mathcal{H} =   \int \frac{d^3k}{l^{-3}} \, \left( \frac{ P_{\mathbf{k}}^2}{2l^{-1}} \,+\, \frac{l^{-1}\omega_{\mathbf{k}}^2 Q_{\mathbf{k}}^2}{2} \right) \,=\, \int \frac{d^3k}{l^{-3}} \, \left( \frac{p_{\mathbf{k}}^2}{2l^{-1}} \,+\, \frac{l^{-1}\omega_{\mathbf{k}}^2 q_{\mathbf{k}}^2}{2} \,+\,  \beta \frac{p_{\mathbf{k}}^4}{3l^{-1}}\right) \,=\, \nonumber \\  \frac{1}{2}  \int d^3 k \, \left(\omega_{\mathbf{k}}  \left[ b^+(\mathbf{k})b(\mathbf{k}) + b(\mathbf{k})b^+(\mathbf{k}) \right] \,+\,  \frac{ \omega_{\mathbf{k}}^2 \beta}{6 \,l^4} \left[b^+(\mathbf{k}) - b(\mathbf{k})\right]^4 \right)   ~.\label{mlsqhambermanmaz} \end{eqnarray} Thus, at the second quantization level the modification amounts to the replacement of Hamiltonian \eqref{desctherulseforsmlq} with Eq.\eqref{mlsqhambermanmaz}. The implementation of modification that arises at the first quantization level into Eq.\eqref{mlsqhambermanmaz} is now straightforward. Namely, one should replace $\omega_{\mathbf{k}}$ with modified relation that comes out of the modified dispersion relation, Eq.\eqref{moddisperrel1}.

The correction term in the Hamiltonian \eqref{mlsqhambermanmaz} is controlled by the dimensionless parameter $\beta /l^2$ and therefore the validity of our approximation demands the smallness of this parameter. For the correction term involves the length scale $l$, the identification of this scale is a subtle question. In view of Eqs.(\ref{mlqmonedim}, \ref{gur}) the deviation from the standard
quantization becomes appreciable at high energies. This fact naturally suggests to identify $l^{-1}$ with the characteristic energy scale of the problem under consideration (for more details see \cite{Berger:2010pj}). For the problem we want to study $l$ is naturally set by the Hubble radius $H^{-1}$.

Now we are in a position to study the minimum length deformed quantization of a free field on the de Sitter background.

\section{Minimum-length deformed quantization of a free field in de Sitter space  }

\noindent Let us consider a masslees neutral scalar field in the FLRW background (we will closely follow the paper \cite{Polarski:1995jg}) 

\[ ds^2 = a^2(
\tau) (d\tau^2 - d\mathbf{x}^2)~, \] \[\mathcal{A}[\Phi] = \frac{1}{2} \int d\tau d^3x \, a^2(\tau) \left[ (\partial_{\tau}\Phi)^2 - \partial_{\mathbf{x}}\Phi \partial_{\mathbf{x}}\Phi  \right] ~.\] To get rid of the overall time-dependent factor, a new field $\varPhi = a(\tau) \Phi$ is usually introduced 

\begin{equation}\label{action}\mathcal{A}[\varPhi] = \frac{1}{2} \int d\tau d^3x  \left[ (\partial_{\tau}\varPhi)^2 - \partial_{\mathbf{x}}\varPhi \partial_{\mathbf{x}}\varPhi    - 2\, \frac{\partial_{\tau} a}{a} \,\varPhi\partial_{\tau}\varPhi    +  \left(\frac{\varPhi\partial_{\tau} a}{a} \right)^2 \right] ~.\nonumber\end{equation} For the Hamiltonian one finds, $\varPi(\mathbf{x}) = \partial_{\tau}\varPhi - \varPhi(\partial_{\tau}a)/a $, 

\begin{equation}\label{hamiltonian} \mathcal{H} = \frac{1}{2}\int d^3x \left[ \varPi^2 + \partial_{\mathbf{x}} \varPhi\partial_{\mathbf{x}} \varPhi + \frac{\partial_{\tau}a}{a}\,(\varPi\varPhi + \varPhi\varPi ) \right]~. \end{equation}

\noindent Expanding $\varPi$ and $\varPhi$ in Fourier modes 

\[ \varPhi(\mathbf{x}) = \int \frac{d^3k}{(2\pi)^{3/2}} \, \varphi (\mathbf{k})\, e^{i\mathbf{k}\mathbf{x}}~,~~~~  \varPi(\mathbf{x}) = \int \frac{d^3k}{(2\pi)^{3/2}} \, \pi (\mathbf{k})\, e^{i\mathbf{k}\mathbf{x}}~, \] where $\mathbf{k}$ denotes the comoving momentum, the Hamiltonian \eqref{hamiltonian} takes the form

\[ \mathcal{H} = \frac{1}{2} \int d^3k \, \left[ \pi(\mathbf{k})\pi^+(\mathbf{k}) + \mathbf{k}^2 \varphi(\mathbf{k})\varphi^+(\mathbf{k})  + \frac{\partial_{\tau}a}{a}\left[\pi(\mathbf{k})\varphi^+(\mathbf{k}) +  \varphi(\mathbf{k}) \pi^+(\mathbf{k}) \right]\right] ~. \]

\noindent  The quantization conditions 
\begin{eqnarray} \left[\varPhi(\mathbf{x}),\, \varPi(\mathbf{y}) \right] = i\delta (\mathbf{x} - \mathbf{y})~,~~\left[\varPhi(\mathbf{x}),\, \varPhi(\mathbf{y}) \right] =0 ~,~~  \left[\varPi(\mathbf{x}),\, \varPi(\mathbf{y}) \right] = 0~,  \nonumber\end{eqnarray} for Fourier amplitudes imply

\begin{eqnarray} \left[\varphi(\mathbf{k}),\, \pi(\mathbf{k'})\right] = i\,\delta(\mathbf{k} + \mathbf{k'})~,~~\left[\varphi(\mathbf{k}),\, \varphi(\mathbf{k'})\right] = 0 ~,~~ \left[\pi(\mathbf{k}),\, \pi(\mathbf{k'})\right] = 0~. \nonumber \end{eqnarray} Defining 

\begin{eqnarray} b(\mathbf{k}) = \frac{1}{\sqrt{2\omega_{\mathbf{k}}}} \left[\omega_{\mathbf{k}}\varphi(\mathbf{k}) + i  \pi(\mathbf{k}) \right] ~,~~ b^+(\mathbf{k}) = \frac{1}{\sqrt{2\omega_{\mathbf{k}}}} \left[\omega_{\mathbf{k}}\varphi(-\mathbf{k}) - i  \pi(-\mathbf{k}) \right]~,\nonumber \end{eqnarray} where $\omega_{\mathbf{k}} = \sqrt{\mathbf{k}^2 }$, one finds 

\begin{eqnarray} \left[b(\mathbf{k}),\,b^+(\mathbf{k'}) \right] = \delta(\mathbf{k} - \mathbf{k'})~,~~ \left[b(\mathbf{k}),\,b(\mathbf{k'}) \right] =0 ~,~~ \left[b^+(\mathbf{k}),\,b^+(\mathbf{k'}) \right] =0~. \nonumber \label{standardcreanni}\end{eqnarray} In terms of $b(\mathbf{k}),\,b^+(\mathbf{k})$ operators the field and momentum operators take the form 

\begin{eqnarray} \varPhi(\mathbf{x}) = \int \frac{d^3k}{\sqrt{(2\pi)^3 2 \omega_{\mathbf{k}} } } \, \left[ b(\mathbf{k})e^{i\mathbf{k}\mathbf{x}} + b^+(\mathbf{k}) e^{-i\mathbf{k} \mathbf{x}} \right]  \nonumber ~, ~~
\varPi(\mathbf{x}) = i \int \frac{ d^3k}{ (2\pi)^{3/2} } \sqrt{ \frac{ \omega_{\mathbf{k}}} {2 }}\,  \left[   b^+(\mathbf{k})e^{-i\mathbf{k}\mathbf{x}} -  b(\mathbf{k})e^{i\mathbf{k}\mathbf{x}} \right]  \nonumber ~, \end{eqnarray} and the Hamiltonian reduces to  

\begin{eqnarray} \mathcal{H} &=& \frac{1}{2}  \int d^3 k \, \left[\omega_{\mathbf{k}}  \left[ b^+(\mathbf{k})b(\mathbf{k}) + b(\mathbf{k})b^+(\mathbf{k}) \right] + i\frac{\partial_{\tau}a}{a} \left[b^+(-\mathbf{k})b^+(\mathbf{k}) - b(\mathbf{k})b(-\mathbf{k}) \right] \right] ~. \nonumber \end{eqnarray} Restricting our attention to the de Sitter space \[ ds^2 = dt^2 -e^{2Ht} d\mathbf{x}^2 ~, ~~~~~~ \tau = - \frac{e^{-Ht}}{H} ~,~~~~~~a(\tau) = - \frac{1}{H\tau}~, \] the Hamiltonian takes the form 

\begin{eqnarray}\label{desitterham} \mathcal{H} &=& \frac{1}{2}  \int d^3 k \, \left[\omega_{\mathbf{k}}  \left[ b^+(\mathbf{k})b(\mathbf{k}) + b(\mathbf{k})b^+(\mathbf{k}) \right] + \frac{i}{\tau} \left[b(\mathbf{k})b(-\mathbf{k}) - b^+(-\mathbf{k})b^+(\mathbf{k})\right] \right] ~.  \end{eqnarray} Heisenberg equation of motion

\begin{equation} \frac{db(\mathbf{k})}{d\tau} = i \left[\mathcal{H},\, b(\mathbf{k}) \right] = -i \omega_{\mathbf{k}}b(\mathbf{k}) - \frac{b^+(-\mathbf{k})}{\tau}~, \nonumber \end{equation} has a well known solution

\begin{equation}\label{bunchdaviesvacuum} b(\tau,\,\mathbf{k}) \,=\,  e^{-i\omega_{\mathbf{k}}\tau } \left(1 - \frac{i}{2\omega_{\mathbf{k}}\tau}\right) b(-\infty,\,\mathbf{k}) + i\,\frac{e^{i\omega_{\mathbf{k}}\tau }}{2\omega_{\mathbf{k}}\tau}\,b^+(-\infty,\,-\mathbf{k}) ~,\end{equation} which for the field operator yields

\begin{equation}\label{desitterfieldoperator}\varPhi(\tau,\,\mathbf{x}) = \int \frac{d^3k}{\sqrt{(2\pi)^3 2 \omega_{\mathbf{k}} } } \, \left[b(-\infty,\,\mathbf{k}) e^{-i(\omega_{\mathbf{k}}\tau - \mathbf{k}\mathbf{x})} \left(1 - \frac{i}{\omega_{\mathbf{k}}\tau}\right)  + b^+(-\infty,\,\mathbf{k}) e^{i(\omega_{\mathbf{k}}\tau - \mathbf{k}\mathbf{x})} \left(1 + \frac{i}{\omega_{\mathbf{k}}\tau}\right) \right] ~. \end{equation}

\noindent From the Hamiltonian Eq.\eqref{desitterham} one sees that as $\tau \rightarrow -\infty$ the deviation from Minkowski space becomes less and less appreciable. This feature allows us to define an unique vacuum state known as the Bunch-Davies vacuum $b(-\infty,\,\mathbf{k})|0\rangle = 0$.

\noindent  Introducing real variables (see subsection "Modification at the second quantization level" in Introduction)

\begin{eqnarray}  Q_{\mathbf{k}} =\, \sqrt{\frac{H^2}{  2 \omega_{\mathbf{k}} }} \left[ b(\mathbf{k}) + b^+(\mathbf{k})\right]~,~~ P_{\mathbf{k}} = \, i  \sqrt{\frac{H^4 \omega_{\mathbf{k}}}{2}} \left[b^+(\mathbf{k}) - b(\mathbf{k})\right] ~,\nonumber \end{eqnarray} the Hamiltonian takes the form

\begin{equation}\label{oscillsumsqueezing} \mathcal{H} =   \int \frac{d^3k}{H^3} \, \left[\left( \frac{ P_{\mathbf{k}}^2}{2H} + \frac{H\omega_{\mathbf{k}}^2 Q_{\mathbf{k}}^2}{2} \right) - \frac{1}{2\tau}\left(  Q_{\mathbf{k}}P_{-\mathbf{k}} + P_{\mathbf{k}}Q_{-\mathbf{k}} \right) \right]~.\end{equation} 

\noindent Assuming now $P_{\mathbf{k}},\, Q_{\mathbf{k}}$ are deformed with respect to the Eq.\eqref{kempfmangano}, the Hamiltonian \eqref{oscillsumsqueezing} takes the form

\begin{equation}\label{mlcorrectedoscillsum} \mathcal{H} =   \int \frac{d^3k}{H^3} \, \left[\left( \frac{ p_{\mathbf{k}}^2}{2H} + \frac{H\omega_{\mathbf{k}}^2 q_{\mathbf{k}}^2}{2} \right) - \frac{1}{2\tau}\left(  q_{\mathbf{k}}p_{-\mathbf{k}} + p_{\mathbf{k}}q_{-\mathbf{k}} \right) +\frac{\beta}{3H}p_{\mathbf{k}}^4 - \frac{\beta}{6\tau}\left(  q_{\mathbf{k}}p_{-\mathbf{k}}^3 + p_{\mathbf{k}}^3q_{-\mathbf{k}} \right) \right] ~.\end{equation}

\noindent In terms of ladder operators this Hamiltonian reads

\begin{eqnarray}\label{mlcorrectedh} && \mathcal{H} = \frac{1}{2}  \int d^3 k \, \left[\omega_{\mathbf{k}}  \left[ b^+(\mathbf{k})b(\mathbf{k}) + b(\mathbf{k})b^+(\mathbf{k})\right] + \frac{i}{\tau} \left[b(\mathbf{k})b(-\mathbf{k}) - b^+(-\mathbf{k})b^+(\mathbf{k})\right] \right. + \frac{\beta H^4 \omega_{\mathbf{k}}^2}{6} \left[b^+(\mathbf{k}) - b(\mathbf{k})\right]^4\nonumber \\
&& \left. +\frac{i\beta H^4 \omega_{\mathbf{k}}}{12\, \tau} \left( \left[ b(\mathbf{k}) + b^+(\mathbf{k})\right] \left[b^+(-\mathbf{k}) - b(-\mathbf{k})\right]^3 + \left[b^+(\mathbf{k}) - b(\mathbf{k})\right]^3\left[ b(-\mathbf{k}) + b^+(-\mathbf{k})\right]   \right) \right] ~.  \end{eqnarray} 

\noindent The Heisenberg equation of motion for $b(\mathbf{k})$ reads now 

\begin{eqnarray}\label{correctedheq} &&\frac{db(\mathbf{k})}{d\tau} = i \left[\mathcal{H},\, b(\mathbf{k}) \right] = -i \omega_{\mathbf{k}}b(\mathbf{k}) \,-\, \frac{b^+(-\mathbf{k})}{\tau} \,- \,\frac{i \beta H^4 \omega_{\mathbf{k}}^2}{12} \left[b^+(\mathbf{k}) - b(\mathbf{k})\right]^3  \\&&  +\, \frac{\beta H^4 \omega_{\mathbf{k}}}{24\, \tau} \left(2 \left[b^+(-\mathbf{k}) - b(-\mathbf{k})\right]^3 + 3\left[b^+(\mathbf{k}) - b(\mathbf{k})\right]^2\left[ b(-\mathbf{k}) + b^+(-\mathbf{k})\right] + 3\left[ b(-\mathbf{k}) + b^+(-\mathbf{k})\right] \left[b^+(\mathbf{k}) - b(\mathbf{k})\right]^2 \right)~.  \nonumber \end{eqnarray}

\noindent Searching for the solution to the first order in $\beta$ in the form 

\[ b(\mathbf{k}) = f(\mathbf{k}) + \beta g(\mathbf{k}) ~,\] from Eq.\eqref{correctedheq} one finds 

\begin{eqnarray}\label{eq1eq2}
&&\frac{df(\mathbf{k})}{d\tau} = -i \omega_{\mathbf{k}}f(\mathbf{k}) \,-\, \frac{f^+(-\mathbf{k})}{\tau} ~, \nonumber \\
&&\frac{dg(\mathbf{k})}{d\tau} = -i \omega_{\mathbf{k}}g(\mathbf{k}) \,-\, \frac{g^+(-\mathbf{k})}{\tau} \,- \,\frac{i  H^4 \omega_{\mathbf{k}}^2}{12} \left[f^+(\mathbf{k}) - f(\mathbf{k})\right]^3  \\&&  +\, \frac{ H^4 \omega_{\mathbf{k}}}{24\, \tau} \left(2 \left[f^+(-\mathbf{k}) - f(-\mathbf{k})\right]^3 + 3\left[f^+(\mathbf{k}) - f(\mathbf{k})\right]^2\left[f(-\mathbf{k}) + f^+(-\mathbf{k})\right] + 3\left[f(-\mathbf{k}) + f^+(-\mathbf{k})\right] \left[f^+(\mathbf{k}) - f(\mathbf{k})\right]^2 \right)~. 
\nonumber \end{eqnarray}

\noindent After solving the first equation of \eqref{eq1eq2}, 

\begin{equation} f(\tau,\,\mathbf{k}) \,=\,  e^{-i\omega_{\mathbf{k}}\tau } \left(1 - \frac{i}{2\omega_{\mathbf{k}}\tau}\right) f(-\infty,\,\mathbf{k}) + i\,\frac{e^{i\omega_{\mathbf{k}}\tau }}{2\omega_{\mathbf{k}}\tau}\,f^+(-\infty,\,-\mathbf{k}) ~,\nonumber \end{equation} and substituting it in the second equation, the system to be solved takes the form 

\begin{equation}\label{system}
\left(\begin{array}{c}
dg(\mathbf{k})/d\tau  \\
dg^+(-\mathbf{k})/d\tau  

\end{array}\right) = \left(\begin{array}{ll}
-i \omega_{\mathbf{k}} & \, -\tau^{-1} \\

-\tau^{-1}  & \,i \omega_{\mathbf{k}}
\end{array}\right) 
\left(\begin{array}{c}
g(\mathbf{k})  \\
g^+(-\mathbf{k})  

\end{array}\right) +  \left(\begin{array}{c}
G(\tau,\, \mathbf{k}) \\
G^+(\tau,\, -\mathbf{k})  

\end{array}\right) ~,\end{equation} where $G(\tau,\, \mathbf{k})$ is given by

\begin{eqnarray}&& G(\tau,\, \mathbf{k}) \,=\, -\,\frac{i  H^4 \omega_{\mathbf{k}}^2}{12} \left[f^+(\tau,\,\mathbf{k}) - f(\tau,\,\mathbf{k})\right]^3  +\, \frac{ H^4 \omega_{\mathbf{k}}}{24\, \tau} \left(2 \left[f^+(\tau,\,-\mathbf{k}) - f(\tau,\,-\mathbf{k})\right]^3 \right. \nonumber \\&&  \left. +\, 3\left[f^+(\tau,\,\mathbf{k}) - f(\tau,\,\mathbf{k})\right]^2\left[f(\tau,\,-\mathbf{k}) + f^+(\tau,\,-\mathbf{k})\right] + 3\left[f(\tau,\,-\mathbf{k}) + f^+(\tau,\,-\mathbf{k})\right] \left[f^+(\tau,\,\mathbf{k}) - f(\tau,\,\mathbf{k})\right]^2 \right)~. \nonumber \end{eqnarray} The solution of the system \eqref{system} is

\begin{eqnarray}\label{syssol} &&
\left(\begin{array}{c}
g(\tau,\,\mathbf{k})  \\
g^+(\tau,\,-\mathbf{k})  

\end{array}\right) = \left(\begin{array}{cc}
e^{-i\omega_{\mathbf{k}}\tau } \left(1 - \frac{i}{2\omega_{\mathbf{k}}\tau}\right) & \, i\,\frac{e^{i\omega_{\mathbf{k}}\tau }}{2\omega_{\mathbf{k}}\tau} \\

-i\,\frac{e^{-i\omega_{\mathbf{k}}\tau }}{2\omega_{\mathbf{k}}\tau}  & \,e^{i\omega_{\mathbf{k}}\tau } \left(1 + \frac{i}{2\omega_{\mathbf{k}}\tau}\right)
\end{array}\right)  \nonumber  \\ && \times \left[  \int\limits_{-\infty}^{\tau}d\xi \left(\begin{array}{cc}
e^{-i\omega_{\mathbf{k}}\xi } \left(1 - \frac{i}{2\omega_{\mathbf{k}}\xi}\right) & \, i\,\frac{e^{i\omega_{\mathbf{k}}\xi }}{2\omega_{\mathbf{k}}\xi} \\

-i\,\frac{e^{-i\omega_{\mathbf{k}}\xi }}{2\omega_{\mathbf{k}}\xi}  & \,e^{i\omega_{\mathbf{k}}\xi } \left(1 + \frac{i}{2\omega_{\mathbf{k}}\xi}\right)
\end{array}\right)^{-1}\left(\begin{array}{c}
G(\xi,\, \mathbf{k}) \\
G^+(\xi,\, -\mathbf{k})  

\end{array}\right)
 \,+\,  \left(\begin{array}{c}
g(-\infty,\,\mathbf{k})  \\
g^+(-\infty,\,-\mathbf{k})  

\end{array}\right)\right]~.\end{eqnarray}

\noindent Thus, for the minimum-length corrected creation and annihilation operators to the first order in $\beta$ one can write

\begin{eqnarray}\label{cranops} &&
\left(\begin{array}{c}
b(\tau,\,\mathbf{k})  \\
b^+(\tau,\,-\mathbf{k})  

\end{array}\right) = \left(\begin{array}{cc}
e^{-i\omega_{\mathbf{k}}\tau } \left(1 - \frac{i}{2\omega_{\mathbf{k}}\tau}\right) & \, i\,\frac{e^{i\omega_{\mathbf{k}}\tau }}{2\omega_{\mathbf{k}}\tau} \\

-i\,\frac{e^{-i\omega_{\mathbf{k}}\tau }}{2\omega_{\mathbf{k}}\tau}  & \,e^{i\omega_{\mathbf{k}}\tau } \left(1 + \frac{i}{2\omega_{\mathbf{k}}\tau}\right)
\end{array}\right)  \nonumber  \\ && \times \left[ \beta \int\limits_{-\infty}^{\tau}d\xi \left(\begin{array}{cc}
e^{i\omega_{\mathbf{k}}\xi } \left(1 + \frac{i}{2\omega_{\mathbf{k}}\xi}\right) & \, -i\,\frac{e^{i\omega_{\mathbf{k}}\xi }}{2\omega_{\mathbf{k}}\xi} \\
i\,\frac{e^{-i\omega_{\mathbf{k}}\xi }}{2\omega_{\mathbf{k}}\xi}  & \, e^{-i\omega_{\mathbf{k}}\xi } \left(1 - \frac{i}{2\omega_{\mathbf{k}}\xi}\right)
\end{array}\right)\left(\begin{array}{c}
G(\xi,\, \mathbf{k}) \\
G^+(\xi,\, -\mathbf{k})  

\end{array}\right)
 \,+\,  \left(\begin{array}{c}
b(-\infty,\,\mathbf{k})  \\
b^+(-\infty,\,-\mathbf{k})  

\end{array}\right)\right]~,\end{eqnarray} where now in $G(\xi,\, \mathbf{k})$ the following replacement is made \begin{equation} f(\xi,\,\mathbf{k}) \,\rightarrow\,  e^{-i\omega_{\mathbf{k}}\xi } \left(1 - \frac{i}{2\omega_{\mathbf{k}}\xi}\right) b(-\infty,\,\mathbf{k}) + i\,\frac{e^{i\omega_{\mathbf{k}}\xi }}{2\omega_{\mathbf{k}}\xi}\,b^+(-\infty,\,-\mathbf{k}) ~.\nonumber \end{equation}

\noindent The field operator takes the form $\varPhi_0 + \beta \varPhi_1$, with $\varPhi_0$ given by Eq.\eqref{desitterfieldoperator} and 

\begin{eqnarray}\label{fieldcorrection}&& \varPhi_1(\tau,\,\mathbf{k}) = \int \frac{d^3k}{\sqrt{(2\pi)^3 2 \omega_{\mathbf{k}} } } \,\int\limits_{-\infty}^{\tau}d\xi \left\{ \left[ \left(e^{-i\omega_{\mathbf{k}}(\tau - \xi) } \left(1 - \frac{i}{2\omega_{\mathbf{k}}\tau}\right) \left(1 + \frac{i}{2\omega_{\mathbf{k}}\xi}\right) - \frac{e^{i\omega_{\mathbf{k}}(\tau - \xi) }}{4\omega_{\mathbf{k}}^2\tau\xi}  \right)G(\xi,\, \mathbf{k})\, - \right. \right.\nonumber \\ && \left. i \left(\frac{e^{-i\omega_{\mathbf{k}}(\tau - \xi)}}{2\omega_{\mathbf{k}}\xi} \left(1 - \frac{i}{2\omega_{\mathbf{k}}\tau}\right) -  \frac{e^{i\omega_{\mathbf{k}}(\tau - \xi)}}{2\omega_{\mathbf{k}}\tau} \left(1 - \frac{i}{2\omega_{\mathbf{k}}\xi}\right)\right)G^+(\xi,\, -\mathbf{k}) \right] e^{i\mathbf{k}\mathbf{x}} \,+ \nonumber \\&& \left[ \left(e^{i\omega_{\mathbf{k}}(\tau - \xi) } \left(1 + \frac{i}{2\omega_{\mathbf{k}}\tau}\right) \left(1 - \frac{i}{2\omega_{\mathbf{k}}\xi}\right) - \frac{e^{-i\omega_{\mathbf{k}}(\tau - \xi) }}{4\omega_{\mathbf{k}}^2\tau\xi}  \right)G^+(\xi,\, \mathbf{k})\, + \right. \nonumber \\ && \left.\left. i \left(\frac{e^{i\omega_{\mathbf{k}}(\tau - \xi)}}{2\omega_{\mathbf{k}}\xi} \left(1 + \frac{i}{2\omega_{\mathbf{k}}\tau}\right) -  \frac{e^{-i\omega_{\mathbf{k}}(\tau - \xi)}}{2\omega_{\mathbf{k}}\tau} \left(1 + \frac{i}{2\omega_{\mathbf{k}}\xi}\right)\right)G(\xi,\, -\mathbf{k}) \right] e^{-i\mathbf{k}\mathbf{x}} \right\}\,\,= \int \frac{d^3k}{\sqrt{(2\pi)^3 2 \omega_{\mathbf{k}} } } \,\int\limits_{-\infty}^{\tau}d\xi \,\times \nonumber \\ &&  \left\{\left[ e^{-i\omega_{\mathbf{k}}(\tau - \xi) } \left(1 + \frac{i}{2\omega_{\mathbf{k}}\xi}\right) - \frac{i e^{-i\omega_{\mathbf{k}}(\tau - \xi)}}{\omega_{\mathbf{k}}\tau} \left(1 + \frac{i}{2\omega_{\mathbf{k}}\xi}\right) - \frac{e^{i\omega_{\mathbf{k}}(\tau - \xi) }}{2\omega_{\mathbf{k}}^2\tau\xi} +\frac{ie^{i\omega_{\mathbf{k}}(\tau - \xi)}}{2\omega_{\mathbf{k}}\xi}  \right]G(\xi,\, \mathbf{k})e^{i\mathbf{k}\mathbf{x}} \,+  \right.\nonumber\\ &&\left. \left[ e^{i\omega_{\mathbf{k}}(\tau - \xi) } \left(1 - \frac{i}{2\omega_{\mathbf{k}}\xi}\right) - \frac{e^{-i\omega_{\mathbf{k}}(\tau - \xi) }}{2\omega_{\mathbf{k}}^2\tau\xi} - \frac{ie^{-i\omega_{\mathbf{k}}(\tau - \xi)}}{2\omega_{\mathbf{k}}\xi} + \frac{ie^{i\omega_{\mathbf{k}}(\tau - \xi)}}{\omega_{\mathbf{k}}\tau} \left(1 - \frac{i}{2\omega_{\mathbf{k}}\xi}\right) \right]G^+(\xi,\, \mathbf{k})e^{-i\mathbf{k}\mathbf{x}} \right\} ~.  \end{eqnarray} For the two point spatial correlation function $\langle 0|\varPhi(\tau,\,\mathbf{x})\varPhi(\tau,\,\mathbf{y})|0\rangle$ (to the first order in $\beta$) one finds   

\begin{equation}\label{twopointcorelation} \langle 0|\varPhi(\tau,\,\mathbf{x})\varPhi(\tau,\,\mathbf{y})|0\rangle = \langle 0|\varPhi_0(\tau,\,\mathbf{x})\varPhi_0(\tau,\,\mathbf{y})|0\rangle  + \beta \langle 0|\varPhi_0(\tau,\,\mathbf{x})\varPhi_1(\tau,\,\mathbf{y})|0\rangle + \beta \langle 0|\varPhi_1(\tau,\,\mathbf{x})\varPhi_0(\tau,\,\mathbf{y})|0\rangle  ~.\end{equation} For both $\varPhi_0$ and $\varPhi_1$ are real fields $\langle 0|\varPhi_0(\tau,\,\mathbf{x})\varPhi_1(\tau,\,\mathbf{y})|0\rangle^* = \langle 0|\varPhi_1^+(\tau,\,\mathbf{x})\varPhi_0^+(\tau,\,\mathbf{y})|0\rangle = \langle 0|\varPhi_1(\tau,\,\mathbf{x})\varPhi_0(\tau,\,\mathbf{y})|0\rangle$. Therefore one gets 

\begin{equation} \langle 0|\varPhi(\tau,\,\mathbf{x})\varPhi(\tau,\,\mathbf{y})|0\rangle = \langle 0|\varPhi_0(\tau,\,\mathbf{x})\varPhi_0(\tau,\,\mathbf{y})|0\rangle  + 2\beta\, \Re \langle 0|\varPhi_0(\tau,\,\mathbf{x})\varPhi_1(\tau,\,\mathbf{y})|0\rangle ~.\end{equation}

\noindent To the end of inflation, that is $\tau \rightarrow 0^-$, one finds (see Appendix)

\begin{eqnarray}
 && \left.\langle 0|\varPhi_0(\tau,\,\mathbf{x})\varPhi_1(\tau,\,\mathbf{y})|0\rangle\right|_{\tau \rightarrow 0^-} \,\rightarrow\, \frac{\delta(\mathbf{0}) H^4}{64\pi^3 \tau^4}\int \frac{d^3k}{\omega_{\mathbf{k}}^4}\left[\frac{ \sin(\mathbf{k}\mathbf{x})\sin(\mathbf{k}\mathbf{y}) }{3} \,+\,  \frac{ \cos(\mathbf{k}\mathbf{x}) \cos(\mathbf{k}\mathbf{y})  }{4} \right]  ~.
\end{eqnarray}

\noindent This sort of $\delta(\mathbf{0})$ appears also when transition probability is calculated from scattering amplitude in QFT \cite{Bogolyubov:1980nc}. In the similar manner we replace it by the three-volume $V$

\[ \delta (\mathbf{k}) \,=\, \frac{1}{(2\pi)^3} \int d^3 x \, e^{i\mathbf{k}\mathbf{x}} ~~\rightarrow ~~ \delta (\mathbf{0}) \,=\, \frac{1}{(2\pi)^3} \int d^3 x  \,=\, \frac{V}{(2\pi)^3}~,\] and set this volume (in view of the problem under consideration) to $H^{-3}$. 

Now returning to the initial field $\Phi = a^{-1}(\tau)\varPhi$ and denoting by $\tau_{end}$ the end of inflation, one finds

\begin{eqnarray}\label{twopointcorrfuncblah}
 &&\left.  2\beta\, \Re\langle 0|\Phi_0(\tau,\,\mathbf{x})\Phi_1(\tau,\,\mathbf{y})|0\rangle\right|_{\tau_{end}} \,\rightarrow\, \frac{\beta \, H^3 }{256 \, \pi^6 \tau_{end}^2}\int \frac{d^3k}{\omega_{\mathbf{k}}^4}\left[\frac{ \sin(\mathbf{k}\mathbf{x})\sin(\mathbf{k}\mathbf{y}) }{3} \,+\,  \frac{ \cos(\mathbf{k}\mathbf{x}) \cos(\mathbf{k}\mathbf{y})  }{4} \right]  ~.
\end{eqnarray} The integral in Eq.\eqref{twopointcorrfuncblah} diverges in the infrared. To find an appropriate IR cutoff let us estimate the size of perturbations to the end of inflation  

\begin{eqnarray}\label{pertsizeestimateendinf}
 &&\left.  \langle 0|\Phi^2(\tau,\,\mathbf{x})|0\rangle\right|_{\tau_{end}} \, \simeq \, \frac{H^2}{(2\pi)^2}\int\limits_{0}^{\infty}\frac{dk}{k} \,+\, \frac{\beta \, H^3 }{256 \, \pi^6 \tau_{end}^2}\int \frac{d^3k}{\omega_{\mathbf{k}}^4}\left[\frac{1}{4} \,+\,  \frac{\sin^2(\mathbf{k}\mathbf{x})}{12}  \right]  ~,
\end{eqnarray} that after omitting the $\sin^2(\mathbf{k}\mathbf{x})$ term in Eq.\eqref{pertsizeestimateendinf} takes the form 

\begin{eqnarray}\label{pertampsquare}
 &&\left.\langle 0|\Phi^2(\tau,\,\mathbf{x})|0\rangle\right|_{\tau_{end}} \, \simeq \, \frac{H^2}{(2\pi)^2} \left[\int\limits_{0}^{\infty}\frac{dk}{k} \,+\, \frac{\beta \, H }{64 \, \pi^3 \tau_{end}^2 }\int\limits_{0}^{\infty} \frac{dk}{k^2}    \right]  ~.
\end{eqnarray} The first integral in Eq.\eqref{pertampsquare} diverges logarithmically. The second integral in Eq.\eqref{pertampsquare} diverges linearly in the infrared. A simple natural way for estimating the scale of an infrared cutoff for the second integral is a follows. The amplitude of perturbations in de Sitter space is proportional to the Gibbons-Hawking temperature $T_{GH} = H/2\pi$ \cite{Linde:2005ht}. The correction to the Gibbons-Hawking temperature due to minimum-length deformed quantization takes the form \cite{Berger:2010pj}  

\begin{equation}\label{corrtogibbhaw} T_{GH} \rightarrow T_{GH} + \beta T^3_{GH} ~.\end{equation} Therefore, for the corrected perturbation amplitude one obtains

\begin{equation}\label{ircutoff} \left.\langle 0|\Phi^2(\tau,\,\mathbf{x})|0\rangle\right|_{\tau_{end}} \, \simeq \,  \frac{H^2}{(2\pi)^2} \left[ 1 \,+\, (\mbox{factor of order unity}) \times \beta H^2 \right]~.\end{equation} By comparing Eq.\eqref{pertampsquare} with Eq.\eqref{ircutoff} one infers that the infrared cutoff should be around $\sim \tau_{end}^{-2}/H$ (or somewhat smaller $\sim 10^{-3}\tau_{end}^{-2}/H$ as $64 \, \pi^3 \approx 2000$). 

It is not out of place to notice that in the case of massive field obeying $m^2 \ll H^2$ one finds \cite{Linde:2005ht}    

\begin{equation} \left.\langle 0|\Phi_0^2(\tau,\,\mathbf{x})|0\rangle\right|_{\tau_{end}} \, \simeq \, \frac{T_{GH}^4}{m^2}~. \label{pertampmassive}\end{equation} Therefore, in view of Eq.\eqref{corrtogibbhaw} one would expect the following correction to Eq.\eqref{pertampmassive} 

\[ \left.\langle 0|\Phi^2(\tau,\,\mathbf{x})|0\rangle\right|_{\tau_{end}} \, \simeq \, \frac{T_{GH}^4}{m^2}\left[1 \,+\, (\mbox{factor of order unity}) \times \beta T_{GH}^2 \right]~. \]

\section{Non-Gaussianities }

It is clear that the non-linear dependence of the field correction on creation and annihilation operators, Eqs.(\ref{fieldcorrection}, \ref{gfunctionofxik}), results in the non-Gaussianity for inflaton perturbations. The non-Gaussianity means that higher order correlation functions can not be expressed completely in terms of the two point function \cite{Bartolo:2004if}. The lowest order non-Gaussian measures are bispectrum (three point correlation function) and trispectrum (four point correlation function). In the case of non-Gaussianity the four point function          

\begin{eqnarray}  \langle 0|\varPhi(\tau,\,\mathbf{x}_1)\varPhi(\tau,\,\mathbf{x}_2)\varPhi(\tau,\,\mathbf{x}_3)\varPhi(\tau,\,\mathbf{x}_4)|0\rangle &=&  \langle 0|\varPhi(\tau,\,\mathbf{x}_1)\varPhi(\tau,\,\mathbf{x}_2)|0\rangle \langle 0| \varPhi(\tau,\,\mathbf{x}_3)\varPhi(\tau,\,\mathbf{x}_4)|0\rangle \,+\, \mbox{permutations} \,+\,  \nonumber \\ && \langle 0|\varPhi(\tau,\,\mathbf{x}_1)\varPhi(\tau,\,\mathbf{x}_2)\varPhi(\tau,\,\mathbf{x}_3)\varPhi(\tau,\,\mathbf{x}_4)|0\rangle_{\mbox{connected}} ~,  \nonumber \end{eqnarray} contains the connected diagrams (the last term) that vanish in the case of Gaussian perturbations. It is easy to see that \begin{eqnarray} \langle 0|\varPhi(\tau,\,\mathbf{x}_1)\varPhi(\tau,\,\mathbf{x}_2)\varPhi(\tau,\,\mathbf{x}_3)\varPhi(\tau,\,\mathbf{x}_4)|0\rangle_{\mbox{connected}} &=& \beta 
\langle 0|\varPhi_1(\tau,\,\mathbf{x}_1)\varPhi_0(\tau,\,\mathbf{x}_2)\varPhi_0(\tau,\,\mathbf{x}_3)\varPhi_0(\tau,\,\mathbf{x}_4)|0\rangle_{\mbox{connected}} \,+ \nonumber \\ && \mbox{permutations with respect to}~ (\mathbf{x}_1,\,\mathbf{x}_2,\,\mathbf{x}_3,\,\mathbf{x}_4)~. \nonumber \end{eqnarray} In our case the field correction $\varPhi_1$ has the following structure, Eqs.(\ref{fieldcorrection}, \ref{gfunctionofxik}),  

\[ \varPhi_1(\mathbf{x})  \,=\, B^3(\mathbf{x}) \,+\, C^2(\mathbf{x})D(\mathbf{x})~.\] So, using the Wick's theorem \cite{Bogolyubov:1980nc}, in the four point function to the first order in $\beta$ we find the following connected diagrams 

\begin{eqnarray} && \langle 0|B(\tau,\,\mathbf{x}_1)\varPhi_0(\tau,\,\mathbf{x}_2)|0\rangle  \langle 0|B(\tau,\,\mathbf{x}_1)\varPhi_0(\tau,\,\mathbf{x}_3)|0\rangle  \langle 0|B(\tau,\,\mathbf{x}_1)\varPhi_0(\tau,\,\mathbf{x}_4)|0\rangle ~,\nonumber \\ && \langle 0|C(\tau,\,\mathbf{x}_1)\varPhi_0(\tau,\,\mathbf{x}_2)|0\rangle  \langle 0|C(\tau,\,\mathbf{x}_1)\varPhi_0(\tau,\,\mathbf{x}_3)|0\rangle  \langle 0|D(\tau,\,\mathbf{x}_1)\varPhi_0(\tau,\,\mathbf{x}_4)|0\rangle ~. \nonumber\end{eqnarray} (It is worth noticing that the three point function vanishes because in this case the vacuum average is taken over an odd number of creation and annihilations operators). Now let us do an explicit calculation. Let us first take the term $\left[b^+(\xi,\,-\mathbf{k}) - b(\xi,\,-\mathbf{k})\right]^3$ from Eq.\eqref{gfunctionofxik}. In Eq.\eqref{fieldcorrection} it is accompanied with $\exp(i\mathbf{k}\mathbf{x})$. We will have the following connected diagram   

\begin{eqnarray}&& e^{i\mathbf{k}\mathbf{x}_1}\langle 0|\left[b^+(\xi,\,-\mathbf{k}) - b(\xi,\,-\mathbf{k})\right]\varPhi_0(\tau,\,\mathbf{x}_2)|0\rangle  \langle 0|\left[b^+(\xi,\,-\mathbf{k}) - b(\xi,\,-\mathbf{k})\right]\varPhi_0(\tau,\,\mathbf{x}_3)|0\rangle  \, \times \nonumber \\ && ~~~~~~~~~~~~~~~~~~~~~~~~~~~~~~~~~~~~~\langle 0|\left[b^+(\xi,\,-\mathbf{k}) - b(\xi,\,-\mathbf{k})\right]\varPhi_0(\tau,\,\mathbf{x}_4)|0\rangle ~, \end{eqnarray} where the entering vacuum averages can be estimated as 

\begin{eqnarray}&& \langle 0|\left[b^+(\xi,\,-\mathbf{k}) - b(\xi,\,-\mathbf{k})\right]\varPhi_0(\tau,\,\mathbf{x}_2)|0\rangle \,=\, \nonumber \\&&  - i\,\frac{e^{-i\omega_{\mathbf{k}}\xi }}{2\omega_{\mathbf{k}}\xi} \, \frac{e^{i(\omega_{\mathbf{k}}\tau - \mathbf{k}\mathbf{x}_2)}} {\sqrt{(2\pi)^3 2 \omega_{\mathbf{k}}}} \left(1 + \frac{i}{\omega_{\mathbf{k}}\tau}\right)  \,-\, e^{-i\omega_{\mathbf{k}}\xi } \left(1 - \frac{i}{2\omega_{\mathbf{k}}\xi}\right) \frac{e^{i(\omega_{\mathbf{k}}\tau + \mathbf{k}\mathbf{x}_2)}} {\sqrt{(2\pi)^3 2 \omega_{\mathbf{k}}}} \left(1 + \frac{i}{\omega_{\mathbf{k}}\tau}\right) ~.\nonumber \end{eqnarray} To estimate the four point function to the end of inflation we may use the following asymptotic expression (see Appendix) 

\begin{eqnarray}\label{asymptgauss} \left.\langle 0|\left[b^+(\xi,\,-\mathbf{k}) - b(\xi,\,-\mathbf{k})\right]\varPhi_0(\tau,\,\mathbf{x}_2)|0\rangle\right|_{\tau \rightarrow 0^-} \,\,\rightarrow \,\,    - \,\frac{i}{\omega^2_{\mathbf{k}}\xi\,\tau} \, \frac{\sin(\mathbf{k}\mathbf{x}_2)} {\sqrt{(2\pi)^3 2 \omega_{\mathbf{k}}}} ~.\end{eqnarray}

Thus, using the asymptotic form of $\mathcal{U}_{\mathbf{k}}(\xi)$, Eq.\eqref{asymexpofing}, and Eq.\eqref{asymptgauss} one finds 

\begin{eqnarray}\label{nongausscont11}&& \Re \int \frac{d^3k}{\sqrt{(2\pi)^3 2 \omega_{\mathbf{k}} } } \,\int\limits_{-\infty}^{\tau}d\xi  \left[ \mathcal{U}_{\mathbf{k}}(\xi)  \,\frac{H^4\omega_{\mathbf{k}}e^{i\mathbf{k}\mathbf{x}_1}}{12\xi}\,\langle 0|\left[b^+(\xi,\,-\mathbf{k}) - b(\xi,\,-\mathbf{k})\right]\varPhi_0(\tau,\,\mathbf{x}_2)|0\rangle   \, \times  \right. \nonumber \\ &&  \left.\left. \langle 0|\left[b^+(\xi,\,-\mathbf{k}) - b(\xi,\,-\mathbf{k})\right]\varPhi_0(\tau,\,\mathbf{x}_3)|0\rangle  \langle 0|\left[b^+(\xi,\,-\mathbf{k}) - b(\xi,\,-\mathbf{k})\right]\varPhi_0(\tau,\,\mathbf{x}_4)|0\rangle  \right] \right|_{\tau \rightarrow 0^-} \, \rightarrow \, \nonumber \\ && \frac{H^4}{96 (2\pi)^6 \tau^6 }\int d^3k \, \frac{\sin(\mathbf{k}\mathbf{x}_1)\sin(\mathbf{k}\mathbf{x}_2)\sin(\mathbf{k}\mathbf{x}_3)\sin(\mathbf{k}\mathbf{x}_4)}{ \omega_{\mathbf{k}}^7 } ~. \end{eqnarray} The other connected diagram contributing to the four point function is (see Eq.\eqref{gfunctionofxik})

\begin{eqnarray}&& e^{i\mathbf{k}\mathbf{x}_1}\langle 0|\left[b^+(\xi,\,\mathbf{k}) - b(\xi,\,\mathbf{k})\right]\varPhi_0(\tau,\,\mathbf{x}_2)|0\rangle  \langle 0|\left[b^+(\xi,\,\mathbf{k}) - b(\xi,\,\mathbf{k})\right]\varPhi_0(\tau,\,\mathbf{x}_3)|0\rangle  \, \times \nonumber \\ && ~~~~~~~~~~~~~~~~~~~~~~~~~~~~~~~~~~~~~\langle 0|\left[b^+(\xi,\,-\mathbf{k}) + b(\xi,\,-\mathbf{k})\right]\varPhi_0(\tau,\,\mathbf{x}_4)|0\rangle ~. \end{eqnarray} One finds

\begin{eqnarray}&& \langle 0|\left[b^+(\xi,\,\mathbf{k}) - b(\xi,\,\mathbf{k})\right]\varPhi_0(\tau,\,\mathbf{x}_2)|0\rangle \,=\, \nonumber \\&&  - i\,\frac{e^{-i\omega_{\mathbf{k}}\xi }}{2\omega_{\mathbf{k}}\xi} \, \frac{e^{i(\omega_{\mathbf{k}}\tau + \mathbf{k}\mathbf{x}_2)}} {\sqrt{(2\pi)^3 2 \omega_{\mathbf{k}}}} \left(1 + \frac{i}{\omega_{\mathbf{k}}\tau}\right)  \,-\, e^{-i\omega_{\mathbf{k}}\xi } \left(1 - \frac{i}{2\omega_{\mathbf{k}}\xi}\right) \frac{e^{i(\omega_{\mathbf{k}}\tau - \mathbf{k}\mathbf{x}_2)}} {\sqrt{(2\pi)^3 2 \omega_{\mathbf{k}}}} \left(1 + \frac{i}{\omega_{\mathbf{k}}\tau}\right) ~,\nonumber \\&& \langle 0|\left[b^+(\xi,\,-\mathbf{k}) + b(\xi,\,-\mathbf{k})\right]\varPhi_0(\tau,\,\mathbf{x}_4)|0\rangle \,=\, \nonumber \\&&  - i\,\frac{e^{-i\omega_{\mathbf{k}}\xi }}{2\omega_{\mathbf{k}}\xi} \, \frac{e^{i(\omega_{\mathbf{k}}\tau - \mathbf{k}\mathbf{x}_4)}} {\sqrt{(2\pi)^3 2 \omega_{\mathbf{k}}}} \left(1 + \frac{i}{\omega_{\mathbf{k}}\tau}\right)  \,+\, e^{-i\omega_{\mathbf{k}}\xi } \left(1 - \frac{i}{2\omega_{\mathbf{k}}\xi}\right) \frac{e^{i(\omega_{\mathbf{k}}\tau + \mathbf{k}\mathbf{x}_4)}} {\sqrt{(2\pi)^3 2 \omega_{\mathbf{k}}}} \left(1 + \frac{i}{\omega_{\mathbf{k}}\tau}\right) ~.\nonumber \end{eqnarray} So, to the end of inflation one may use the following expressions (see Appendix)

\begin{eqnarray} \label{asymptgauss11}&& \left.\langle 0|\left[b^+(\xi,\,\mathbf{k}) - b(\xi,\,\mathbf{k})\right]\varPhi_0(\tau,\,\mathbf{x}_2)|0\rangle\right|_{\tau \rightarrow 0^-} \,\,\rightarrow \,\,  \frac{i}{\omega_{\mathbf{k}}^2\xi \,\tau} \, \frac{\sin(\mathbf{k}\mathbf{x}_2)}{\sqrt{(2\pi)^3 2 \omega_{\mathbf{k}}}} ~,\\\label{asymptgauss22}&& \left.\langle 0|\left[b^+(\xi,\,-\mathbf{k}) + b(\xi,\,-\mathbf{k})\right]\varPhi_0(\tau,\,\mathbf{x}_4)|0\rangle\right|_{\tau \rightarrow 0^-} \,\,\rightarrow \,\, \frac{1}{\omega_{\mathbf{k}}^2\xi \,\tau} \, \frac{\cos(\mathbf{k}\mathbf{x}_4)}{\sqrt{(2\pi)^3 2 \omega_{\mathbf{k}}}} ~.\end{eqnarray}

Again, using the asymptotic form of $\mathcal{U}_{\mathbf{k}}(\xi)$, Eq.\eqref{asymexpofing}, and Eqs.(\ref{asymptgauss11}, \ref{asymptgauss22}) one finds 

\begin{eqnarray}\label{nongausscont22}&& \Re \int \frac{d^3k}{\sqrt{(2\pi)^3 2 \omega_{\mathbf{k}} } } \,\int\limits_{-\infty}^{\tau}d\xi  \left[ \mathcal{U}_{\mathbf{k}}(\xi)  \,\frac{H^4\omega_{\mathbf{k}}e^{i\mathbf{k}\mathbf{x}_1}}{8\xi}\, \langle 0|\left[b^+(\xi,\,\mathbf{k}) - b(\xi,\,\mathbf{k})\right]\varPhi_0(\tau,\,\mathbf{x}_2)|0\rangle   \, \times  \right. \nonumber \\ &&  \left.\left. \langle 0|\left[b^+(\xi,\,\mathbf{k}) - b(\xi,\,\mathbf{k})\right]\varPhi_0(\tau,\,\mathbf{x}_3)|0\rangle \langle 0|\left[b^+(\xi,\,-\mathbf{k}) + b(\xi,\,-\mathbf{k})\right]\varPhi_0(\tau,\,\mathbf{x}_4)|0\rangle  \right] \right|_{\tau \rightarrow 0^-} \, \rightarrow \, \nonumber \\ && \frac{H^4}{64 (2\pi)^6 \tau^6 }\int d^3k \, \frac{\cos(\mathbf{k}\mathbf{x}_1)\sin(\mathbf{k}\mathbf{x}_2)\sin(\mathbf{k}\mathbf{x}_3)\cos(\mathbf{k}\mathbf{x}_4)}{ \omega_{\mathbf{k}}^7 } ~. \end{eqnarray} The reason we have taken only real parts in Eqs.(\ref{nongausscont11}, \ref{nongausscont22}) is that after using the complete expression of $\varPhi_1$, see Eq.\eqref{compactformofphy1}, only the real parts will survive (and in addition will be multiplied by $2$).   

Now returning to the initial field $\Phi = a^{-1}(\tau)\varPhi$ and using the infrared cutoff introduced in the previous section, one finds the following magnitude for non-Gaussianity to the first order in $\beta$

\begin{eqnarray}\label{nongaussianityfinalresult} && \langle 0|\Phi(\tau_{end},\,\mathbf{x}_1)\Phi(\tau_{end},\,\mathbf{x}_2)\Phi(\tau_{end},\,\mathbf{x}_3)\Phi(\tau_{end},\,\mathbf{x}_4)|0\rangle_{\mbox{connected}}   \,\,\propto \,\,  \nonumber \\&& \frac{\beta H^8}{48 (2\pi)^6 \tau^2_{end} }\int\limits_{\tau_{end}^{-2}/H} \frac{d^3k}{\omega_{\mathbf{k}}^7} \,\left[ \sin(\mathbf{k}\mathbf{x}_1)\sin(\mathbf{k}\mathbf{x}_2)\sin(\mathbf{k}\mathbf{x}_3)\sin(\mathbf{k}\mathbf{x}_4) \,+\, \frac{2}{3}\,\cos(\mathbf{k}\mathbf{x}_1)\sin(\mathbf{k}\mathbf{x}_2)\sin(\mathbf{k}\mathbf{x}_3)\cos(\mathbf{k}\mathbf{x}_4) \right] ~. \nonumber \end{eqnarray} Estimating order of magnitude for non-Gaussianity (we use $\tau_{end}^{-2} = H^2e^{2N}$, where $N$ stands for the number of $e$-foldings) 

\begin{equation} \mbox{Non-Gaussianity} \,\, \simeq \,\,  \frac{\beta H^8}{48 (2\pi)^6 \tau^2_{end} }\int\limits_{\tau_{end}^{-2}/H}^{\infty} \frac{d^3k}{k^7} \,\,=\,\, \frac{\pi \beta H^{12}\tau_{end}^6}{48 (2\pi)^6 } \,\,=\,\, \frac{\pi \beta H^{6}e^{-6N}}{48 (2\pi)^6 }~,  \nonumber \end{equation} one infers that, taking into account that for a successful inflation $N \simeq 70$, the non-Gaussianity becomes suppressed by the factor $e^{-420} \approx 10^{-182}$.

\section{Concluding remarks}

In the introduction we tried to clearly outline a scheme for implementation of minimum-length deformed quantization in the QFT formalism at the level of second quantization. This scheme is contrasted with the implementation of minimum-length deformed prescription in the first quantization picture. Minimum-length deformed quantum mechanics combined with the field theory at the second quantization level reveals qualitatively new features as compared to such combination at the first quantization level \cite{Mania:2009dy, Berger:2010pj}, and may teach us something deeper about what is in fact implied by this sort of quantization.

An analytic solution for the field operator is found to the first order in deformation parameter on de Sitter background. That is interesting in its own right and can be used for working out of various phenomenological questions. The field operator shows up a non-linear dependence on creation and annihilation operators and therefore leads to the non-Gaussianity for the perturbation field in the cosmological context.     

Similar solution for the Minkowskian background was found in \cite{Berger:2010pj}. It is instructive to notice that this result applied to the black hole radiation yields the correct corrections to the black hole entropy \cite{Berger:2010pj}. 

Using the solution for field operator, the two point correlation function has been estimated. The correction to the two point correlation function shows up an infrared divergence (typical for massless theories). Hence, one is compelled to invoke some physical arguments to pin down the infrared cutoff. For this purpose we use the results of paper \cite{Berger:2010pj} that enables one to estimate the order of magnitude for the correction and correspondingly gives an idea about the cutoff. The correction to the two point function is of the same order as obtained previously in the framework of modified field theory at the first quantization level due to minimum length deformed prescription, see for instance most resent paper about it \cite{Palma:2008tx}.    

Non-Gaussianity appears at the lowest level in four point function (the three point function merely vanishes). The magnitude of non-Gaussianity is strongly suppressed by the factor $e^{-6N}$, where $N$ stands for the number of $e$-foldings. 

It would be interesting to generalize present discussion to the massive field.

\vspace{0.5cm}

{\bf Acknowledgements:~~}  Author is indebted to Micheal S. Berger, Jens C. Niemeyer and Zurab K. Silagadze for useful comments.  

\section*{Appendix}

For notational convenience instead of $b(-\infty,\,\mathbf{k}),\,b^+(-\infty,\,\mathbf{k})$ we simply use $b(\mathbf{k}),\,b^+(\mathbf{k})$; besides that we denote minimum-length corrected operators by tilde and keep the notations $b(\xi,\,\mathbf{k}),\,b^+(\xi,\,\mathbf{k})$ for 

\begin{equation} b(\xi,\,\mathbf{k}) \,=\,  e^{-i\omega_{\mathbf{k}}\xi } \left(1 - \frac{i}{2\omega_{\mathbf{k}}\xi}\right) b(\mathbf{k}) + i\,\frac{e^{i\omega_{\mathbf{k}}\xi }}{2\omega_{\mathbf{k}}\xi}\,b^+(-\mathbf{k}) ~,~~b^+(\xi,\,\mathbf{k}) \,=\,  e^{i\omega_{\mathbf{k}}\xi } \left(1 + \frac{i}{2\omega_{\mathbf{k}}\xi}\right) b^+(\mathbf{k}) - i\,\frac{e^{-i\omega_{\mathbf{k}}\xi }}{2\omega_{\mathbf{k}}\xi}\,b(-\mathbf{k}) ~,\end{equation} see Eq.\eqref{bunchdaviesvacuum}. For calculating the corrections to the correlation function \eqref{twopointcorelation}, first we have to estimate the following vacuum averages $\langle 0|b(\mathbf{k}')G(\xi,\, \mathbf{k})|0\rangle$  and $\langle 0|b(\mathbf{k}')G^+(\xi,\, \mathbf{k})|0\rangle$, where

\begin{eqnarray}\label{gfunctionofxik}&& G(\xi,\, \mathbf{k}) \,=\, -\,\frac{i  H^4 \omega_{\mathbf{k}}^2}{12} \left[b^+(\xi,\,\mathbf{k}) - b(\xi,\,\mathbf{k})\right]^3  +\, \frac{ H^4 \omega_{\mathbf{k}}}{24\, \xi} \left(2 \left[b^+(\xi,\,-\mathbf{k}) - b(\xi,\,-\mathbf{k})\right]^3 \right. \nonumber \\&&  \left. +\, 3\left[b^+(\xi,\,\mathbf{k}) - b(\xi,\,\mathbf{k})\right]^2\left[b(\xi,\,-\mathbf{k}) + b^+(\xi,\,-\mathbf{k})\right] + 3\left[b(\xi,\,-\mathbf{k}) + b^+(\xi,\,-\mathbf{k})\right] \left[b^+(\xi,\,\mathbf{k}) - b(\xi,\,\mathbf{k})\right]^2 \right)~.  \end{eqnarray} Splitting Eq.\eqref{gfunctionofxik} 
\begin{eqnarray} G(\xi,\, \mathbf{k}) &=& G_1(\xi,\, \mathbf{k}) \,+\, G_2(\xi,\, \mathbf{k})~, \nonumber \\
 G_1(\xi,\, \mathbf{k}) &=& -\,\frac{i  H^4 \omega_{\mathbf{k}}^2}{12} \left[b^+(\xi,\,\mathbf{k}) - b(\xi,\,\mathbf{k})\right]^3  +\, \frac{ H^4 \omega_{\mathbf{k}}}{12\, \xi}  \left[b^+(\xi,\,-\mathbf{k}) - b(\xi,\,-\mathbf{k})\right]^3  ~,\nonumber \\  G_2(\xi,\, \mathbf{k}) &=&   \frac{ H^4 \omega_{\mathbf{k}}}{8\, \xi} \left( \left[b^+(\xi,\,\mathbf{k}) - b(\xi,\,\mathbf{k})\right]^2\left[b(\xi,\,-\mathbf{k}) + b^+(\xi,\,-\mathbf{k})\right] + \left[b(\xi,\,-\mathbf{k}) + b^+(\xi,\,-\mathbf{k})\right] \left[b^+(\xi,\,\mathbf{k}) - b(\xi,\,\mathbf{k})\right]^2 \right)~,\nonumber \end{eqnarray} one sees that $G_1^+(\xi,\, \mathbf{k}) = - G_1(\xi,\, \mathbf{k})$ and $G_2^+(\xi,\, \mathbf{k}) = G_2(\xi,\, \mathbf{k})$. So, calculating $\langle 0|b(\mathbf{k}')G(\xi,\, \mathbf{k})|0\rangle$, we easily estimate $\langle 0|b(\mathbf{k}')G^+(\xi,\, \mathbf{k})|0\rangle$ as well.

Non-zero contributions to the $\langle 0|b(\mathbf{k}')\left[b^+(\xi,\,\mathbf{k}) - b(\xi,\,\mathbf{k})\right]^3|0\rangle$ come from the terms containing equal number of creation and annihilation operators that cancel each other \cite{Bogolyubov:1980nc}

\begin{eqnarray}&& i\,\frac{e^{i\omega_{\mathbf{k}}\xi }}{\left(2\omega_{\mathbf{k}}\xi\right)^3}\,\langle 0|b(\mathbf{k}')b(-\mathbf{k})b^+(-\mathbf{k}) b^+(-\mathbf{k})|0\rangle \,=\, 2i\, \delta(\mathbf{k}' + \mathbf{k})\delta(\mathbf{0})\, \frac{e^{i\omega_{\mathbf{k}}\xi }}{\left(2\omega_{\mathbf{k}}\xi\right)^3}~,\nonumber\\ && i\,\frac{e^{i\omega_{\mathbf{k}}\xi }}{\left(2\omega_{\mathbf{k}}\xi\right)^3}\, \langle 0|b(\mathbf{k}')b^+(-\mathbf{k})b(-\mathbf{k}) b^+(-\mathbf{k})|0\rangle \,=\, i\, \delta(\mathbf{k}' + \mathbf{k})\delta(\mathbf{0})\,\frac{e^{i\omega_{\mathbf{k}}\xi }}{\left(2\omega_{\mathbf{k}}\xi\right)^3}\, ~,\nonumber\\ && i\,\frac{e^{i\omega_{\mathbf{k}}\xi }}{2\omega_{\mathbf{k}}\xi} \left(1 + \frac{1}{\left(2\omega_{\mathbf{k}}\xi\right)^2}\right) \langle 0|b(\mathbf{k}')b(\mathbf{k})b^+(\mathbf{k}) b^+(-\mathbf{k})|0\rangle \,=\, i\,\delta(\mathbf{k}' + \mathbf{k})\delta(\mathbf{0}) \,\frac{e^{i\omega_{\mathbf{k}}\xi }}{2\omega_{\mathbf{k}}\xi} \left(1 + \frac{1}{\left(2\omega_{\mathbf{k}}\xi\right)^2}\right)~,\nonumber\\&& -\frac{e^{i\omega_{\mathbf{k}}\xi }}{\left(2\omega_{\mathbf{k}}\xi\right)^2} \left(1 + \frac{i}{2\omega_{\mathbf{k}}\xi}\right) \langle 0|b(\mathbf{k}')b(-\mathbf{k})b^+(-\mathbf{k}) b^+(\mathbf{k})|0\rangle \,=\, -\,\delta(\mathbf{k}' - \mathbf{k})\delta(\mathbf{0})\, \frac{e^{i\omega_{\mathbf{k}}\xi }}{\left(2\omega_{\mathbf{k}}\xi\right)^2} \left(1 + \frac{i}{2\omega_{\mathbf{k}}\xi}\right) ~,\nonumber\\&& -\,e^{i\omega_{\mathbf{k}}\xi } \left(1 + \frac{1}{\left(2\omega_{\mathbf{k}}\xi\right)^2}\right)\left(1 + \frac{i}{2\omega_{\mathbf{k}}\xi}\right) \langle 0|b(\mathbf{k}')b(\mathbf{k}) b^+(\mathbf{k})b^+(\mathbf{k})|0\rangle = -2 \delta(\mathbf{k}' - \mathbf{k})\delta(\mathbf{0})\,e^{i\omega_{\mathbf{k}}\xi } \left(1 + \frac{1}{\left(2\omega_{\mathbf{k}}\xi\right)^2}\right)\left(1 + \frac{i}{2\omega_{\mathbf{k}}\xi}\right) ~, \nonumber \\&& -\,e^{i\omega_{\mathbf{k}}\xi } \left(1 + \frac{1}{\left(2\omega_{\mathbf{k}}\xi\right)^2}\right)\left(1 + \frac{i}{2\omega_{\mathbf{k}}\xi}\right) \langle 0|b(\mathbf{k}') b^+(\mathbf{k})b(\mathbf{k})b^+(\mathbf{k})|0\rangle \,=\,  -\,\delta(\mathbf{k}' - \mathbf{k})\delta(\mathbf{0}) \,e^{i\omega_{\mathbf{k}}\xi } \left(1 + \frac{1}{\left(2\omega_{\mathbf{k}}\xi\right)^2}\right)\left(1 + \frac{i}{2\omega_{\mathbf{k}}\xi}\right)~, \nonumber\end{eqnarray} 

\noindent  where $\delta(\mathbf{0})$ stands for $\delta(\mathbf{0}) \equiv \delta(\mathbf{k} - \mathbf{k})$.

Making replacement $\mathbf{k} \rightarrow -\mathbf{k}$ one gets $\langle 0|b(\mathbf{k}')\left[b^+(\xi,\,\mathbf{k}) - b(\xi,\,\mathbf{k})\right]^3|0\rangle \rightarrow\langle 0|b(\mathbf{k}')\left[b^+(\xi,\,-\mathbf{k}) - b(\xi,\,-\mathbf{k})\right]^3|0\rangle$. So, the above calculations readily determine this term as well.

Now let us estimate $\langle 0|b(\mathbf{k}') \left[b^+(\xi,\,\mathbf{k}) - b(\xi,\,\mathbf{k})\right]^2\left[b(\xi,\,-\mathbf{k}) + b^+(\xi,\,-\mathbf{k})\right] |0\rangle$. Terms giving non-zero contributions to this vacuum average look as follows

\begin{eqnarray}&& -\,\frac{e^{i\omega_{\mathbf{k}}\xi }}{\left(2\omega_{\mathbf{k}}\xi\right)^2}\left(1 + \frac{i}{2\omega_{\mathbf{k}}\xi}\right)\langle 0|b(\mathbf{k}')b(-\mathbf{k})b^+(-\mathbf{k}) b^+(-\mathbf{k})|0\rangle \,=\,  -2\,\delta(\mathbf{k}' + \mathbf{k})\delta(\mathbf{0})\,\frac{e^{i\omega_{\mathbf{k}}\xi }}{\left(2\omega_{\mathbf{k}}\xi\right)^2}\left(1 + \frac{i}{2\omega_{\mathbf{k}}\xi}\right)~,\nonumber\\ && -\,\frac{e^{i\omega_{\mathbf{k}}\xi }}{\left(2\omega_{\mathbf{k}}\xi\right)^2}\left(1 + \frac{i}{2\omega_{\mathbf{k}}\xi}\right) \langle 0|b(\mathbf{k}')b^+(-\mathbf{k})b(-\mathbf{k}) b^+(-\mathbf{k})|0\rangle \,=\, -\, \delta(\mathbf{k}' + \mathbf{k})\delta(\mathbf{0}) \,\frac{e^{i\omega_{\mathbf{k}}\xi }}{\left(2\omega_{\mathbf{k}}\xi\right)^2}\left(1 + \frac{i}{2\omega_{\mathbf{k}}\xi}\right)~,\nonumber\\ && -\,e^{i\omega_{\mathbf{k}}\xi }\left(1 + \frac{1}{\left(2\omega_{\mathbf{k}}\xi\right)^2}\right) \left(1 + \frac{i}{2\omega_{\mathbf{k}}\xi}\right) \langle 0|b(\mathbf{k}')b(\mathbf{k})b^+(\mathbf{k}) b^+(-\mathbf{k})|0\rangle \,=\, -\, \delta(\mathbf{k}' + \mathbf{k})\delta(\mathbf{0})\, e^{i\omega_{\mathbf{k}}\xi }\left(1 + \frac{1}{\left(2\omega_{\mathbf{k}}\xi\right)^2}\right) \left(1 + \frac{i}{2\omega_{\mathbf{k}}\xi}\right)~,\nonumber\\&& - i\,\frac{e^{i\omega_{\mathbf{k}}\xi }}{\left(2\omega_{\mathbf{k}}\xi\right)^3} \langle 0|b(\mathbf{k}')b(-\mathbf{k})b^+(-\mathbf{k}) b^+(\mathbf{k})|0\rangle \,=\, - i\, \delta(\mathbf{k}' - \mathbf{k})\delta(\mathbf{0})\,\frac{e^{i\omega_{\mathbf{k}}\xi }}{\left(2\omega_{\mathbf{k}}\xi\right)^3} ~,\nonumber\\&& -\, i\,\frac{e^{i\omega_{\mathbf{k}}\xi }}{2\omega_{\mathbf{k}}\xi} \left(1 + \frac{1}{\left(2\omega_{\mathbf{k}}\xi\right)^2}\right)\langle 0|b(\mathbf{k}')b(\mathbf{k}) b^+(\mathbf{k})b^+(\mathbf{k})|0\rangle \,=\,  -2i\,\delta(\mathbf{k}' - \mathbf{k})\delta(\mathbf{0})\,\frac{e^{i\omega_{\mathbf{k}}\xi }}{2\omega_{\mathbf{k}}\xi} \left(1 + \frac{1}{\left(2\omega_{\mathbf{k}}\xi\right)^2}\right) ~, \nonumber \\&& -i\, \frac{e^{i\omega_{\mathbf{k}}\xi }}{2\omega_{\mathbf{k}}\xi} \left(1 + \frac{1}{\left(2\omega_{\mathbf{k}}\xi\right)^2}\right)\langle 0|b(\mathbf{k}') b^+(\mathbf{k})b(\mathbf{k})b^+(\mathbf{k})|0\rangle \,=\, -i\, \delta(\mathbf{k}' - \mathbf{k})\delta(\mathbf{0}) \, \frac{e^{i\omega_{\mathbf{k}}\xi }}{2\omega_{\mathbf{k}}\xi} \left(1 + \frac{1}{\left(2\omega_{\mathbf{k}}\xi\right)^2}\right) ~. \nonumber \end{eqnarray} 

Finally we have to estimate the expression $\langle 0| b(\mathbf{k}') \left[b(\xi,\,-\mathbf{k}) + b^+(\xi,\,-\mathbf{k})\right] \left[b^+(\xi,\,\mathbf{k}) - b(\xi,\,\mathbf{k})\right]^2 |0\rangle$. One finds that it contains the following terms

\begin{eqnarray}&& -\,\frac{e^{i\omega_{\mathbf{k}}\xi }}{\left(2\omega_{\mathbf{k}}\xi\right)^2}\left(1 - \frac{i}{2\omega_{\mathbf{k}}\xi}\right) \langle 0|b(\mathbf{k}')b(-\mathbf{k})b^+(-\mathbf{k}) b^+(-\mathbf{k})|0\rangle \,=\, -2\,\delta(\mathbf{k}' + \mathbf{k})\delta(\mathbf{0}) \,\frac{e^{i\omega_{\mathbf{k}}\xi }}{\left(2\omega_{\mathbf{k}}\xi\right)^2}\left(1 - \frac{i}{2\omega_{\mathbf{k}}\xi}\right) ~,\nonumber\\ && -\,\frac{e^{i\omega_{\mathbf{k}}\xi }}{\left(2\omega_{\mathbf{k}}\xi\right)^2}\left(1 - \frac{i}{2\omega_{\mathbf{k}}\xi}\right) \langle 0|b(\mathbf{k}')b^+(-\mathbf{k})b(-\mathbf{k}) b^+(-\mathbf{k})|0\rangle \,=\, -\, \delta(\mathbf{k}' + \mathbf{k})\delta(\mathbf{0}) \,\frac{e^{i\omega_{\mathbf{k}}\xi }}{\left(2\omega_{\mathbf{k}}\xi\right)^2}\left(1 - \frac{i}{2\omega_{\mathbf{k}}\xi}\right) ~,\nonumber\\ && - \,\frac{e^{i\omega_{\mathbf{k}}\xi }}{\left(2\omega_{\mathbf{k}}\xi\right)^2} \left(1 + \frac{i}{2\omega_{\mathbf{k}}\xi}\right)\langle 0|b(\mathbf{k}')b(\mathbf{k})b^+(\mathbf{k}) b^+(-\mathbf{k})|0\rangle \,=\, -\, \delta(\mathbf{k}' + \mathbf{k})\delta(\mathbf{0}) \,\frac{e^{i\omega_{\mathbf{k}}\xi }}{\left(2\omega_{\mathbf{k}}\xi\right)^2} \left(1 + \frac{i}{2\omega_{\mathbf{k}}\xi}\right)~,\nonumber\\&&  -i\,\frac{e^{i\omega_{\mathbf{k}}\xi }}{2\omega_{\mathbf{k}}\xi} \left(1 + \frac{1}{\left(2\omega_{\mathbf{k}}\xi\right)^2}\right)  \langle 0|b(\mathbf{k}')b(-\mathbf{k})b^+(-\mathbf{k}) b^+(\mathbf{k})|0\rangle \,=\, -i\, \delta(\mathbf{k}' - \mathbf{k})\delta(\mathbf{0}) \,\frac{e^{i\omega_{\mathbf{k}}\xi }}{2\omega_{\mathbf{k}}\xi} \left(1 + \frac{1}{\left(2\omega_{\mathbf{k}}\xi\right)^2}\right)~,\nonumber\\&& - i\,\frac{e^{i\omega_{\mathbf{k}}\xi }}{2\omega_{\mathbf{k}}\xi} \left(1 + \frac{i}{2\omega_{\mathbf{k}}\xi}\right)^2  \langle 0|b(\mathbf{k}')b(\mathbf{k}) b^+(\mathbf{k})b^+(\mathbf{k})|0\rangle  \,=\, -i 2\,\delta(\mathbf{k}' - \mathbf{k})\delta(\mathbf{0})\,\frac{e^{i\omega_{\mathbf{k}}\xi }}{2\omega_{\mathbf{k}}\xi} \left(1 + \frac{i}{2\omega_{\mathbf{k}}\xi}\right)^2  ~, \nonumber \\&& -i\,\frac{e^{i\omega_{\mathbf{k}}\xi }}{2\omega_{\mathbf{k}}\xi} \left(1 + \frac{i}{2\omega_{\mathbf{k}}\xi}\right)^2  \langle 0|b(\mathbf{k}') b^+(\mathbf{k})b(\mathbf{k})b^+(\mathbf{k})|0\rangle \,=\,  -i\,\delta(\mathbf{k}' - \mathbf{k})\delta(\mathbf{0})\,\frac{e^{i\omega_{\mathbf{k}}\xi }}{2\omega_{\mathbf{k}}\xi} \left(1 + \frac{i}{2\omega_{\mathbf{k}}\xi}\right)^2  ~. \nonumber \end{eqnarray} 

Writing Eq.\eqref{fieldcorrection} in a compact form \begin{equation}\label{compactformofphy1}\varPhi_1(\tau,\,\mathbf{x}) \,=\, \int \frac{d^3k}{\sqrt{(2\pi)^3 2 \omega_{\mathbf{k}} } } \,\int\limits_{-\infty}^{\tau}d\xi  \left[ \mathcal{U}_{\mathbf{k}}(\xi) G(\xi,\, \mathbf{k})e^{i\mathbf{k}\mathbf{x}} \,+\, \mathcal{U}_{\mathbf{k}}^*(\xi) G^+(\xi,\, \mathbf{k})e^{-i\mathbf{k}\mathbf{x}}  \right] ~,\end{equation} where \[ \mathcal{U}_{\mathbf{k}}(\xi) \,\equiv \,  e^{-i\omega_{\mathbf{k}}(\tau - \xi) } \left(1 + \frac{i}{2\omega_{\mathbf{k}}\xi}\right) - \frac{i e^{-i\omega_{\mathbf{k}}(\tau - \xi)}}{\omega_{\mathbf{k}}\tau} \left(1 + \frac{i}{2\omega_{\mathbf{k}}\xi}\right) - \frac{e^{i\omega_{\mathbf{k}}(\tau - \xi) }}{2\omega_{\mathbf{k}}^2\tau\xi} +\frac{ie^{i\omega_{\mathbf{k}}(\tau - \xi)}}{2\omega_{\mathbf{k}}\xi}   ~,\] the correlation function $\langle 0|\varPhi_0(\tau,\,\mathbf{x})\varPhi_1(\tau,\,\mathbf{y})|0\rangle$ takes the form \begin{eqnarray}\label{finalexpcorfunct} &&\langle 0|\varPhi_0(\tau,\,\mathbf{x})\varPhi_1(\tau,\,\mathbf{y})|0\rangle \,=\,  \int \frac{d^3k\,d^3k'}{2(2\pi)^3 \sqrt{\omega_{\mathbf{k}}\omega_{\mathbf{k}'} } } \left(1 - \frac{i}{\omega_{\mathbf{k}'}\tau}\right)e^{-i(\omega_{\mathbf{k}'}\tau - \mathbf{k}'\mathbf{x})} \,\, \times \nonumber \\ && \int\limits_{-\infty}^{\tau} d\xi  \left[ \mathcal{U}_{\mathbf{k}}(\xi) \langle 0|b(\mathbf{k}')G(\xi,\, \mathbf{k})|0\rangle e^{i\mathbf{k}\mathbf{y}} \,+\, \mathcal{U}_{\mathbf{k}}^*(\xi) \langle 0|b(\mathbf{k}')G^+(\xi,\, \mathbf{k})|0\rangle e^{-i\mathbf{k}\mathbf{y}}  \right] \,= \nonumber \\  && \int \frac{d^3k\,d^3k'}{(2\pi)^3 \sqrt{\omega_{\mathbf{k}}\omega_{\mathbf{k}'} } } \left(1 - \frac{i}{\omega_{\mathbf{k}'}\tau}\right)e^{-i(\omega_{\mathbf{k}'}\tau - \mathbf{k}'\mathbf{x})}  \,\,\times \nonumber \\  && \int\limits_{-\infty}^{\tau}d\xi  \left[ i \langle 0|b(\mathbf{k}')G_1(\xi,\, \mathbf{k})|0\rangle \, \Im \left(\mathcal{U}_{\mathbf{k}}(\xi)e^{i\mathbf{k}\mathbf{y}}\right) \,+\,  \langle 0|b(\mathbf{k}')G_2(\xi,\, \mathbf{k})|0\rangle \, \Re \left( \mathcal{U}_{\mathbf{k}}(\xi) e^{i\mathbf{k}\mathbf{y}}\right)  \right] ~. \end{eqnarray} To evaluate this correlation function to the end of inflation, that is, $\tau \rightarrow 0^-$, one can use an asymptotic expression for the integral over $\xi$. It is easy to notice that in the limit $\tau \rightarrow 0^-$, the main contribution to this integral comes from smaller values of $\xi$, that is, $\xi \rightarrow 0$.  Namely, for a typical term in this integral one finds ($n> 1,\,\tau < 0$)

\[ \left|\int\limits_{-\infty}^{\tau}d\xi \,\frac{e^{i\omega_{\mathbf{k}}\xi }}{\left(\omega_{\mathbf{k}}\xi\right)^n}\right|  \,\leq \, \int\limits_{-\infty}^{\tau}d\xi \,\frac{1}{\left(\omega_{\mathbf{k}}\left|\xi\right|\right)^n} \,=\,  \frac{1}{\omega_{\mathbf{k}}^n(n-1)} \,\, \times \left\{\begin{array}{l} \tau^{1-n} ~~~~~ \mbox{if}~ n~ \mbox{is odd}\,, \\ -\,\tau^{1-n} ~~\mbox{if}~ n~ \mbox{is even}\,. \end{array}\right.\] So, to get a leading term when $\tau \rightarrow 0^-$, we replace        

\begin{eqnarray}\label{asymexpofing}&&\mathcal{U}_{\mathbf{k}}(\xi) \,\rightarrow \,  \frac{\xi}{\tau} ~, \nonumber \\&&\langle 0|b(\mathbf{k}')G_1(\xi,\, \mathbf{k})|0\rangle \,\rightarrow \,  \frac{ H^4 \omega_{\mathbf{k}}}{3\, \xi} \, \frac{i}{\left(2\omega_{\mathbf{k}}\xi\right)^3}  \left[ \delta(\mathbf{k}' - \mathbf{k})  \,-\, \delta(\mathbf{k}' + \mathbf{k}) \right]\delta(\mathbf{0})~,  \\&& \langle 0|b(\mathbf{k}')G_2(\xi,\, \mathbf{k})|0\rangle  \,\rightarrow \, \,-\,\frac{ H^4 \omega_{\mathbf{k}}}{4\, \xi} \frac{i}{\left(2\omega_{\mathbf{k}}\xi\right)^3}  \left[ \delta(\mathbf{k}' - \mathbf{k})  \,+\, \delta(\mathbf{k}' + \mathbf{k}) \right]\delta(\mathbf{0})~.  \nonumber \end{eqnarray} After integrating over $\xi$ one gets 

\begin{eqnarray}\label{intxiasymp}&& \int\limits_{-\infty}^{\tau}d\xi  \left[ i \langle 0|b(\mathbf{k}')G_1(\xi,\, \mathbf{k})|0\rangle \, \Im \left(\mathcal{U}_{\mathbf{k}}(\xi)e^{i\mathbf{k}\mathbf{y}}\right) \,+\,  \langle 0|b(\mathbf{k}')G_2(\xi,\, \mathbf{k})|0\rangle \, \Re \left( \mathcal{U}_{\mathbf{k}}(\xi) e^{i\mathbf{k}\mathbf{y}}\right)  \right]  \rightarrow \nonumber \\&&  \frac{ H^4 \delta(\mathbf{0})}{48\, \omega_{\mathbf{k}}^2\tau^3} \, \left[ \delta(\mathbf{k}' - \mathbf{k})  \,-\, \delta(\mathbf{k}' + \mathbf{k}) \right]\sin(\mathbf{k}\mathbf{y}) \,+\, \frac{i H^4 \delta(\mathbf{0})}{64\, \omega_{\mathbf{k}}^2\tau^3} \, \left[ \delta(\mathbf{k}' - \mathbf{k})  \,+\, \delta(\mathbf{k}' + \mathbf{k}) \right]\cos(\mathbf{k}\mathbf{y})~.\end{eqnarray} Using the result \eqref{intxiasymp} in Eq.\eqref{finalexpcorfunct} one finds

\begin{eqnarray}\label{asymexpoftwopointcor}
 &&\left.\langle 0|\varPhi_0(\tau,\,\mathbf{x})\varPhi_1(\tau,\,\mathbf{y})|0\rangle\right|_{\tau \rightarrow 0^-} \,\rightarrow\, \delta(\mathbf{0}) \int \frac{d^3k}{(2\pi)^3 \omega_{\mathbf{k}}}\left[\frac{ 2H^4 }{48\, \omega_{\mathbf{k}}^3\tau^4} \,\sin(\mathbf{k}\mathbf{x})\sin(\mathbf{k}\mathbf{y}) \,+\,  \frac{ 2H^4 }{64\, \omega_{\mathbf{k}}^3\tau^4} \,\cos(\mathbf{k}\mathbf{x}) \cos(\mathbf{k}\mathbf{y}) \right]  ~.
\end{eqnarray}

\end{document}